\documentclass[10pt]{article}
\usepackage{amsmath, amssymb, slashed, latexsym, cite, tensor, simplewick, bbm}
\usepackage{wrapfig,lipsum,booktabs}
\usepackage{mathtools}
\usepackage[dvipsnames]{xcolor}
\usepackage[a4paper, left=2.5cm, right=2.5cm, top=3.5cm, bottom=3cm]{geometry}
\usepackage[hidelinks]{hyperref}
\hypersetup{
	colorlinks=true,
	linkcolor=blue,
	citecolor=magenta,
}
\usepackage{booktabs,dcolumn,caption}
\newcommand{\im}{\mathrm{i}}

\newcommand{\diff}{\mathrm{d}}
\newcommand{\pa}{\partial}
\newcommand{\tr}{\mathrm{tr}}
\newcommand{\sfrac}[2]{{\textstyle\frac{#1}{#2}}}

\renewcommand{\=}{\ =\ }
\newcommand{\unity}{\mathbbm{1}}

\newcommand{\with}{\quad\textrm{with}\quad}
\renewcommand{\>}{\rangle}
\newcommand{\<}{\langle}

\newcommand{\drm}{\mathrm{d}}
\newcommand{\rR}{r}
\newcommand{\nc}{N}
\newcommand{\rD}{\mathrm{D}}
\newcommand{\FP}{\ensuremath{\Delta_{\mathrm{FP}}[A]}}
\newcommand{\MSS}{\ensuremath{\Delta_{\mathrm{MSS}}[A]}}
\newcommand{\SU}{\mathrm{SU}}
\newcommand{\bC}{\ensuremath{\bar{C}}}
\newcommand{\A}{\widetilde A}
\newcommand{\R}{\widetilde R}
\newcommand{\G}{\mathcal{G}}
\newcommand{\stF}{{}^{\star}\hspace{-2pt}\tF}
\newcommand{\tD}{\ensuremath{\widetilde{D}}}

\newcommand{\trD}{\ensuremath{\widetilde{\rD}}}
\newcommand{\tC}{\ensuremath{\widetilde{C}}}
\newcommand{\tbC}{\ensuremath{\widetilde{\bar{C}}}}
\newcommand{\tF}{\ensuremath{\widetilde{F}}}
\newcommand{\tlambda}{\widetilde{\lambda}}
\newcommand{\Tr}{\ensuremath{\operatorname{tr}}}
\newcommand{\blambda}{\ensuremath{\bar{\lambda}}}
\newcommand{\intdx}{\ensuremath{\int \drm^4 x\;}}

\newcommand{\partialA}{\ensuremath{\partial{\cdot}A}}
\newcommand{\Boxi}{\ensuremath{\Box^{-1}}}
\newcommand{\tA}{\ensuremath{\widetilde{A}}}

\newcommand{\tblambda}{\widetilde{\bar{\lambda}}}

\newcommand{\slA}{\slashed{A}}
\renewcommand{\t}{\times}
\newcommand{\Diff}{\mathrm{D}}

\newcommand{\slp}{\;\;\slashed{\partial}}

\newcommand{\Rl}[1]{\ensuremath{\stackrel{\leftarrow}{R}^{\mathrm{\,#1}}}}

\makeatletter
\newcommand{\pushright}[1]{\ifmeasuring@#1\else\omit\hfill$\displaystyle#1$\fi\ignorespaces}
\newcommand{\pushleft}[1]{\ifmeasuring@#1\else\omit$\displaystyle#1$\hfill\fi\ignorespaces}
\makeatother

\begin{document}
	\pagenumbering{gobble}
	\sloppy
	\title{\bf\huge \vspace{-25pt}An improved Nicolai map\\ for super Yang--Mills theory}
	\date{~}
	
	\author{\vspace{-30pt}\phantom{.}\\[12pt]
		{\scshape\Large Olaf Lechtenfeld \ and \ Maximilian Rupprecht}
		\\[24pt]
		Institut f\"ur Theoretische Physik\\ 
		and\\ Riemann Center for Geometry and Physics
		\\[8pt]
		Leibniz Universit\"at Hannover \\ 
		Appelstra{\ss}e 2, 30167 Hannover, Germany
		\\[24pt]
	} 
	
	\clearpage
	\maketitle
	\thispagestyle{empty}
	
	\begin{abstract}
	\noindent\large
	Adding a topological theta term to the action of ${\cal N}{=}\,1$ $D{=}4$ super Yang--Mills theory modifies its Nicolai map. 
	For the BPS value of the theta angle a chiral version of the map emerges, which allows for a considerable simplification compared to the non-chiral formulation. 
	We exhibit these improvements to all orders in perturbation theory and compute the map to fourth order in the coupling on the Laudau-gauge hypersurface. 
	The second-order contribution vanishes, and antisymmetrizations are more manifest. All checks are verified to third order. 
	\end{abstract}
	
	\newpage
	\pagenumbering{arabic}
	
	\noindent\textbf{Introduction.\ }
	The Nicolai map formalism can be applied to any off-shell supersymmetric field theory (with coupling parameters~$g$). It was originally contrived by Hermann Nicolai \cite{Nic1,Nic2,Nic3} and developed further by Flume, Dietz and one of the authors in the 1980s~\cite{FL,DL1,DL2,L1,L2}. In recent years, it has experienced a sort of renaissance with a number of modern papers~\cite{ANPP,NP,ALMNPP,AMPP,LR1,MN,LR2,LN,R,M,LR3,L3}. The defining property of the Nicolai map, a nonlinear and nonlocal field transformation of the bosonic fields~$\phi\,\mapsto T_g\phi$, is that any correlation function of the interacting theory reduces to a free-field ($g{=}0$) correlator,
	\begin{equation}\label{eq:correlator_property}
		\bigl\langle X[\phi] \bigr\rangle_g \= \bigl\langle X[T_g^{-1}\phi] \bigr\rangle_0
	\end{equation}
	for any functional~$X[\phi]$, where $T_g$ can always be inverted perturbatively near the identity.
	We here specialize to unbroken $\mathcal{N}\=1$ supersymmetric Yang--Mills in the Wess--Zumino gauge in four--dimensional Minkowski spacetime $\mathbb{R}^{1,3}\ni x$ with field content $(A,\lambda,D)$ in the adjoint representation of the gauge group. 
	Choosing a gauge-fixing function $\G(A)$ adds, via the Faddeev--Popov trick and the 't Hooft averaging, a gauge-fixing term
	depending on ghost fields $C$ and~$\bar{C}$ and a gauge parameter~$\xi$ to the action. For convenience, we choose the Landau gauge
	\begin{equation}
		\G(A) \= \pa^\mu\! A_\mu\ ,
	\end{equation}
	in which the map seems to take its simplest form \cite{LR2,MN}. We note that the Nicolai map is constructed to respect any gauge fixing, $\mathcal{G}(T_gA)=\mathcal{G}(A)$.
	For the construction of the map it is essential that the $g$-derivative of the action can be expressed as a supervariation plus a compensating Slavnov variation. This is achieved best in a particular field scaling \cite{LR1}
  	\begin{equation}
  		\A \= g\,A \= \A_\mu\,\diff x^\mu \qquad\textrm{and}\qquad 
		\widetilde{F} \= g\,F \= \diff\A + \A\wedge\A \= \sfrac12 \widetilde{F}_{\mu\nu}\,\diff x^\mu\wedge\diff x^\nu\ .
  	\end{equation}
	The full action, amended by the usual topological theta term, splits up into an invariant piece and a gauge-fixing part,
	\begin{equation}
		\begin{aligned}
			S_{\textrm{\tiny SUSY}}&\= S_{\mathrm{inv}}\ +\ S_{\mathrm{gf}}\ ,\\
			S_{\mathrm{inv}}&\=-\sfrac{1}{g^2}\int\!\diff^4 x\ \tr\bigl\{ \sfrac14 \widetilde{F}_{\mu\nu}\widetilde{F}^{\mu\nu} 
			- \sfrac{g^2 \theta
			  }{32\pi^2}\tF_{\mu\nu}\stF^{\mu\nu}\ +\ \textrm{fermions}\ +\ \textrm{auxiliaries} \bigr\}\ ,\\
			S_{\mathrm{gf}}&\=-\sfrac{1}{g^2}\int\!\diff^4 x\ \tr\bigl\{\sfrac{1}{2\xi}\G(\A)^2\ +\ \textrm{ghosts} \bigr\}\ ,
		\end{aligned}
	\end{equation}
	with the dual field strength
	\begin{equation}
		\stF^{\mu\nu}\=\sfrac12 \epsilon^{\mu\nu\rho\lambda}\tF_{\rho\lambda}\ .
	\end{equation}
	Setting up perturbation theory in the gauge coupling~$g$ will eventually require returning to the untilded variables.
	In the following we fix
	\begin{equation}
		\theta'\ :=\ \sfrac{g^2\theta}{8\pi^2}
	\end{equation}
	to a constant. In other words, we investigate a flow in the $(g,\theta)$ parameter space along lines $\theta=\tfrac{8\pi^2}{g^2}\theta'$ determined by fixed values of~$\theta'$. Therefore, the correlators in \eqref{eq:correlator_property} as well as the Nicolai map implicitly depend on~$\theta'$! In perturbation theory, we expand around the vacuum, where $A$ is pure gauge, and thus may restrict ourselves to the topologically trivial sector in configuration space, where $\int F{\wedge}F=0$. Hence, perturbative correlators cannot depend on~$\theta'$, and we are allowed to dial any complex value for it!\footnote{
		In a nonperturbative treatment, the reality of the action demands $\theta'{\in}\mathbb{R}$. For Euclidean signature, one must take $\theta'{\in}\im\mathbb{R}$.}
	Indeed, there exist two special imaginary values $\theta'=\pm\im$ for which one obtains a chiral formulation of the Nicolai map. This `chiral Nicolai map' for ${\cal N}{=}\,1$ super Yang--Mills theory is the subject of this paper.\footnote{This was already partially explored in \cite{L1} using Weyl spinors instead of the now--preferred Majorana-formulation.} In \cite{LR3}, the possibility of adding a topological theta term to ${\cal N}=1$ supersymmetric quantum mechanics was thoroughly studied. In particular, it was found that for `magical' theta values, the Nicolai map becomes a unique linear function in the coupling $g$. Here, we do not obtain a truncation of the map, but still find significant simplifications in comparison to the construction with $\theta\=0$ \cite{ALMNPP, MN}.

	\noindent\textbf{Conventions and notation.\ }
	We choose the mostly plus metric $\eta^{\mu\nu}=\operatorname{diag}(-,+,+,+)$ and the Clifford algebra
	\begin{equation}
		\bigl\{\gamma^\mu,\gamma^\nu\bigr\}\=2\,\eta^{\mu\nu}\ ,
	\end{equation}
	as well as the definition of the `fifth' gamma matrix
	\begin{equation}
		\gamma^5\=\im\,\gamma^0\gamma^1\gamma^2\gamma^3\ .
	\end{equation}
	We also employ the shorthand
	\begin{equation}
		\gamma^{\mu\nu}\=\sfrac12(\gamma^\mu\gamma^\nu-\gamma^\nu\gamma^\mu)\ .
	\end{equation}
	For simplicity we take the gauge group to be $\SU(\nc)$ with real antisymmetric structure constants $f^{abc}$ such that
	\begin{equation}\label{eq:structure_constants}
		f^{abc}f^{abd}\=\nc\,\delta^{cd}\ ,\qquad a,b,\ldots\=1, 2, \ldots , \nc^2{-}1\ .
	\end{equation}
	The Yang--Mills fields are labeled as $A^a_\mu$ ($\mu=0,1,2,3$), but we often suppress color indices.
	We summarize the quantities that appear in the coupling flow operator in the next section. The fermionic propagator $S$ is the Green's function of the covariant derivative $\rD_\mu\=\partial_\mu+gA_\mu\times\ $ contracted with the gamma matrices (we also suppress Majorana spinor indices $\alpha,\beta,\ldots$),
	\begin{equation}
		S\ =\ \slashed{\rD}^{-1} \= - \bcontraction{}{\lambda}{\blambda}{\ }\lambda\ \blambda\ ,
	\end{equation}
	whereas the ghost propagator in Landau gauge is given by
	\begin{equation}
		G \= \bigl(\partial\cdot\rD\bigr)^{-1}\=-\im\;\bcontraction{}{C}{\bC}{\; }C\ \bC\ .
	\end{equation}
	They can be expanded in the coupling (on the Landau gauge hypersurface $\partial^\mu A_\mu\equiv 0$) as 
	\begin{equation}
		\begin{aligned}
			&S\=\slp C\ -\ g\slp C\slA S\= \slp C\ -\ g\slp C\slA \slp C\ +\ g^2\slp C\slA \slp C\slA \slp C\ -\ \ldots\ ,\\
			&G\= C\ -\ g C A{\cdot}\partial\; G\= C\ -\ g C A{\cdot}\partial\; C\ +\ g^2 C A{\cdot}\partial\; C A{\cdot}\partial\; C\ -\ \ldots\ ,
		\end{aligned}
	\end{equation}
	in terms of the free scalar propagator
	\begin{equation}
		C\=\Boxi\ .
	\end{equation}
	We often adopt from Section~4 of \cite{ALMNPP} (also used in \cite{LR2,R}) the shorthand (de Witt) notation for multiplying quantities in color and position space. This means that all objects are multiplied as color matrices or vectors, and integration kernels are convoluted with insertions of bosonic fields $A_\mu$. For example, we write in two equivalent notations the expression
	\begin{equation}
		\partial^\rho C A^\lambda \partial_\mu C A_\rho{\times} A_\lambda \qquad\Leftrightarrow\qquad 
		\int\!\drm^4y\ \drm^4z\   \partial^\rho C(x{-}y)\; (f^{abc}A^{b\,\lambda})(y)\; \partial_\mu C(y{-}z)\; (f^{cde}A^d_\rho)(z) A_\lambda^e(z)\ .
	\end{equation}
	To make such expressions more compact, we often use the shorthand notation $\partial_\nu C\equiv C_\nu$, $\partial_\mu\partial_\nu C\equiv C_{\mu\nu}$ and so on.
	
	\noindent\textbf{Coupling flow operator.\ }
	With the addition of the topological term, the $g$-derivative of the action is generated by a supervariation ($\delta_\alpha$) up to a Slavnov variation ($s$)
	\begin{equation}
		\pa_g S_{\textrm{\tiny SUSY}} \= -\sfrac{1}{g^3}\,\bigl\{ 
		\delta_\alpha \Delta_\alpha'
		- \sqrt{g}\,s\,\Delta_{\textrm{gh}} \bigr\}
		  \qquad\with\qquad \Delta_{\textrm{gh}} \= \intdx\;\tr\, \bigl\{\tbC\,\G(\A))\bigr\}\ .
	\end{equation}
	The $\theta'$-dependence enters in the superfield component $\Delta_\alpha'$ via
	\begin{equation}\label{eq:modified_Delta}
		\Delta_\alpha'\=\Delta_\beta\ [1+\im\theta'\gamma^5]_{\beta\alpha} \qquad\with\qquad 
		\Delta_\alpha[\tA, \tlambda, \tD] \= \sfrac{1}{4}\intdx\; \tr\,\bigl\{\sfrac{1}{2}\gamma^{\mu\nu} \tlambda \tF_{\mu\nu}\,-\,\im\,\gamma_5 \tlambda \tD\bigr\}_\alpha\ ,
	\end{equation}
	where $\Delta_\alpha$ is a gauge-invariant fermionic functional that generates the invariant part of the action without a topological term, see~\cite{ALMNPP,LR2,MN}. The supervariations are given by 
	\begin{equation}\label{eq:supervariations}
		\delta_\alpha \tA_\nu=-(\tblambda\gamma_\nu)_\alpha\ ,\qquad
		\delta_\alpha \tlambda_\beta=\sfrac{1}{2}(\gamma^{\mu\nu})_{\beta\alpha}\tF_{\mu\nu}+\im\tD(\gamma_5)_{\beta\alpha}\ ,\qquad
		\delta_\alpha \tD=-\im(\trD_\mu \tblambda \gamma_5 \gamma^\mu)_{\alpha} \ ,
	\end{equation}
	and the relevant Slavnov variations by
	\begin{equation}\label{eq:slavnov_variations}
		s\tA_\mu=\sqrt{g}\ \trD_\mu \tC \qquad\textrm{and}\qquad  s\,\tbC=\sfrac{1}{\sqrt{g}}\sfrac{1}{\xi}\;\G(\tA)\ .
	\end{equation}
	One can now employ the usual construction of the Nicolai map via the coupling flow operator $\R[\A]$, that captures the effect of a $g$~derivative on expectation values $\bigl\<X[\A]\bigr\>_g$ (after integrating out
	gaugini, ghosts and auxiliaries) via
	\begin{equation}
		\pa_g \bigl\< X[\A] \bigr\>_g \= \bigl\< \bigl( \pa_g + \sfrac1g\R[\A] \bigr) X[\A] \bigr\>_g\ .
	\end{equation}
	It is a linear functional differential operator given by~\cite{DL1,L1}
	\begin{equation} \label{gaugeR}
		\R[\A] \= -\im\,\bcontraction{}{\Delta}{_\alpha'[\A]\ }{\delta} \Delta_\alpha'[\A]\ \delta_\alpha 
		+\sfrac{\im}{\sqrt{g}}\,\bcontraction{}{\Delta}{_{\textrm{gh}}[\A]\ }{s} \Delta_{\textrm{gh}}[\A]\ s
		-\sfrac{1}{\sqrt{g}}\,\bcontraction{}{\Delta}{_\alpha'[\A]\ \bigl(}{\delta}  \Delta_\alpha'[\A]\ \bigl(\delta_\alpha 
		\bcontraction{}{\Delta}{_{\textrm{gh}}[\A]\bigr)\ }{s} \Delta_{\textrm{gh}}[\A]\bigr)\ s\ ,
	\end{equation}
	where contractions indicate either gaugino or ghost propagators and the auxiliary fields have been integrated out ($\tD=0$). To develop a power series expansion of the Nicolai map \cite{L1,LR1,LR2,MN} we must rescale back $\widetilde{A}=g\,A$ and find the `rescaled flow operator'
	\begin{equation} \label{Rrel}
		R_g[A] \= \sfrac1g\,\bigl( \R[gA] - \smallint\! A\,\sfrac{\delta}{\delta A} \bigr)\ ,
	\end{equation}
	with the only novelty here being the insertion of the square brackets in \eqref{eq:modified_Delta}. Acting to the left, the operator can be written as \\[-8pt]
	\begin{equation}\label{eq:R_g_expanded}
		\stackrel{\leftarrow}{R_g}[A]\=
		-\sfrac{1}{8}\; \stackrel{\leftarrow}{\sfrac{\delta}{\delta A_\mu}}\ \bigl[\delta\indices{_\mu^\nu}-\Diff_\mu G\,\partial^\nu\bigr]\ \Tr\bigl\{\gamma_\nu S \gamma^{\rho\lambda}\bigl[\unity+\im\theta'\gamma^5\bigr]\bigr\}\;A_\rho{\times} A_\lambda\ .
	\end{equation}
	
	\noindent\textbf{Simplifications.\ }
	The Nicolai map can be constructed from the perturbative expansion of the coupling flow operator \\[-12pt]
	\begin{equation}\label{eq:series_R}
		R_g[A] \= \sum_{k=1}^\infty g^{k-1} \rR_k[A] \= \rR_1[A] + g\,\rR_2[A] + g^2 \rR_3[A] + \ldots\ ,
	\end{equation}
	via the `universal formula' \cite{LR1}. The first few orders are given by
	\begin{equation}\label{eq:exp_Tg}
		\begin{aligned}
			T_gA &\= A \ -\ g\,\rR_1 A \ -\ \sfrac12g^2\bigl(\rR_2-\rR_1^2\bigr)A\ -\ 
			\sfrac16g^3\bigl(2\rR_3-2\rR_2\rR_1-\rR_1\rR_2+\rR_1^3\bigr)A \\[4pt]
			&\quad -\sfrac{1}{24}g^4\bigl(6\rR_4-6\rR_3\rR_1-2\rR_1\rR_3+2\rR_1\rR_2\rR_1-3\rR_2\rR_2+3\rR_2\rR_1^2+\rR_1^2\rR_2-\rR_1^4\bigr)A \ +\ {\cal O}(g^5)\ .
		\end{aligned}
	\end{equation}
	We find that a decomposition of the covariant projector
	\begin{equation}\label{eq:cov_projector}
		\delta\indices{^\mu_\nu}-\Diff_\mu(\partial\cdot\Diff)^{-1}\partial^\nu
		\=\underbrace{\delta\indices{_\mu^\nu}}_{\mathrm{inv}}\ \underbrace{-\,\partial_\mu C\partial^\nu}_{\mathrm{lgt}}\ \underbrace{-\,g[A_\mu-C_\mu A\cdot\partial]\,G\,\partial^\nu}_{\mathrm{gh}}\ ,
	\end{equation}  \\[-8pt]
	into an `invariant', `longitudinal' and `ghost' part is very useful in our construction. 
	According to \eqref{eq:cov_projector}, we split up the coupling flow operator \eqref{eq:R_g_expanded} into three contributions, \\[-8pt]
	\begin{equation}
		R_g\=R_g^{\mathrm{inv}}+R_g^{\mathrm{lgt}}+R_g^{\mathrm{gh}}
		\= \sum_{k=1}^{\infty} g^{k-1} (r_k^{\textrm{inv}}+r_k^{\textrm{lgt}}+r_k^{\textrm{gh}})\ .
	\end{equation}  \\[-8pt]
	Introducing the shorthand
	\begin{equation}\label{eq:E}
		E_\mu[A;x]\=\sfrac{1}{8}\;\Tr\bigl\{\gamma_\mu S \gamma^{\rho\lambda}\bigl[\unity+\im\theta'\gamma^5\bigr]\bigr\}\;A_\rho\times A_\lambda
		\=E^{(1)}_\mu+g\,E^{(2)}_\mu+g^2E^{(3)}_\mu+\ldots
	\end{equation}
	and making use of
	\begin{equation}
		\Diff^\nu E_\nu\=0\qquad\iff\qquad \partial^\nu E_\nu\=-g A^\nu \t E_\nu\ ,
	\end{equation}
	the three parts of the coupling flow operator can be written compactly as \\[-4pt]
	\begin{equation}
		\Rl{inv}_g\= - \stackrel{\leftarrow}{\sfrac{\delta}{\delta A_\mu}}E_\mu\ ,\qquad 
		\Rl{lgt}_g\=g\,\stackrel{\leftarrow}{\sfrac{\delta}{\delta A_\mu}}\partial_\mu C A^\nu\t E_\nu\ ,\qquad 
		\Rl{gh}_g\=-\,g^2 \stackrel{\leftarrow}{\sfrac{\delta}{\delta A_\mu}}[A_\mu-C_\mu A\cdot\partial]\,G\,A^\nu\t E_\nu\ .
	\end{equation}
	At ${\cal O}(g^0)$ only the invariant part contributes ($\rR_1=\rR^{\mathrm{inv}}_1$), 
	and at ${\cal O}(g^1)$ there is no ghost contribution ($\rR_2=\rR^{\mathrm{inv}}_2+\rR^{\mathrm{lgt}}_2$).
	Adding the longitudinal to the invariant part, i.e.~$\rR^{\mathrm{inv}}_k \to \rR^{\mathrm{inv+lgt}}_k$, amounts to a simple antisymmetrization, \\[-4pt]
	\begin{equation}
		-\tfrac18 \stackrel{\leftarrow}{\sfrac{\delta}{\delta A_\mu}}
		\Tr\bigl\{\gamma_\mu\gamma_\alpha \ldots  \gamma^{\rho\lambda}\bigl[\ldots\bigr]\bigr\}\,C^\alpha\ldots A_\rho{\times} A_\lambda \quad\ \Rightarrow\ \quad
		-\tfrac18\stackrel{\leftarrow}{\sfrac{\delta}{\delta A_\mu}}
                \Tr\bigl\{\gamma_{\mu\alpha} \ldots  \gamma^{\rho\lambda}\bigl[\ldots\bigr]\bigr\}\,C^\alpha\ldots A_\rho{\times} A_\lambda\ ,
	\end{equation}
	which is automatic for $k{=}1$, consistent with $\rR^{\mathrm{inv}}_1=\rR^{\mathrm{inv+lgt}}_1$.

	{}From now on we specialize to the BPS value $\theta'=-\im$. 
	This implements a chiral projection in the gamma trace, and one may switch to a Weyl-spinor formulation~\cite{FL,DL1,DL2,L1,L2}.
	The decisive advantage for the chiral formulation with $\theta'=\pm\im$ comes from the Fierz identity 
	\begin{equation} \label{Fierz}
		[\gamma^\nu(\unity{+}\gamma^5)]_{\alpha\beta}\ [\gamma_\nu(\unity{-}\gamma^5)]_{\gamma\delta}
		\=-2\,(\unity{-}\gamma^5)_{\alpha\delta}\ (\unity{+}\gamma^5)_{\gamma\beta}\ ,
	\end{equation}
	which implies that
	\begin{equation} \label{fusetrace}
		\rR_{k-1}^{\mathrm{inv}}\;\rR_1\;A\=(\rR_{k}^{\mathrm{inv}}+\rR_{k}^{\mathrm{lgt}})\;A\qquad\textrm{for}\quad \theta'=\pm\im\quad\textrm{and}\quad k\ \geq\ 2\ .
	\end{equation}
	Inserting this into \eqref{eq:exp_Tg}, the second order vanishes entirely because $\rR_1^2 A = \rR_2 A$, and to fourth order we obtain
	\begin{equation}
	\begin{aligned}
	T_gA_\mu &\= A_\mu \ -\ g\;\rR_1A_\mu \ -\ \sfrac13g^3\bigl(\rR_3^{\mathrm{gh}}-\rR_2^{\mathrm{lgt}}\rR_1\bigr)A_\mu\\
	&\hspace{2.6cm} -\ \sfrac{1}{12}g^4\bigl(3\rR_4^{\mathrm{gh}}-3(\rR_3^{\mathrm{lgt}}+\rR_3^{\mathrm{gh}})\,\rR_1
	-\rR_1(\rR_3^{\mathrm{gh}}-\rR_2^{\mathrm{lgt}}\rR_1)\bigr)A_\mu \ +\ {\cal O}(g^5) \\[4pt]
	&\= A_\mu\ +\ \tfrac18 g\;\tr\bigl\{\gamma_{\mu\alpha}\gamma^{\rho\lambda}\bigl[\unity{+}\gamma^5\bigr]\bigr\}\,
	C^\alpha A_\rho{\times}A_\lambda \\
	&\qquad\quad +\ \tfrac1{24}g^3\, [A_\mu-C_\mu A{\cdot}\partial]\,C\,A^\nu\;
	\tr\bigl\{\gamma_\nu\gamma_\beta\gamma^{\rho\lambda}\bigl[\unity{+}\gamma^5\bigr]\bigr\}\,
	C^\beta A_\rho{\times}A_\lambda \\
	&\qquad\quad -\ \tfrac1{96}g^3\,\tr\bigl\{ \gamma_{\mu\alpha}\gamma^{\nu\beta}\bigl[\unity{+}\gamma^5\bigr]\bigr\} \;
	\tr\bigl\{ \gamma_{\sigma\gamma}\gamma^{\rho\lambda}\bigl[\unity{+}\gamma^5\bigr]\bigr\}\,
	C^\alpha A_\nu C_\beta A^\sigma C^\gamma A_\rho{\times}A_\lambda \\
	&\qquad\quad -\ \sfrac{1}{12}g^4\bigl(3\rR_4^{\mathrm{gh}}-3(\rR_3^{\mathrm{lgt}}+\rR_3^{\mathrm{gh}})\,\rR_1
	-\rR_1(\rR_3^{\mathrm{gh}}-\rR_2^{\mathrm{lgt}}\rR_1)\bigr)A_\mu \ +\ {\cal O}(g^5) \ ,
	\end{aligned}
	\end{equation}
	where we spelled out the first three orders explicitly.
	This pattern persists to all orders. In the universal formula \cite{LR1} at order $g^n$ we sum over all compositions ${\bf n}=
	(n_1,n_2,\ldots,n_s)$ with $n=\sum_i n_i$. The form of the Stirling-type coefficients $c_{\bf n}$ (see \cite{LR1}) allows us to pairwise combine terms to
	\begin{equation}
	\begin{aligned}
	T_g A &\= A\ -\ g\,r_1 A\ +\ {\sum_{\bf n}}^\prime g^n\ c_{\bf n}\ r_{n_s}\ldots r_{n_2}\,(r_k - r_{k-1}r_1)\;A  \\
	&\= A\ -\ g\,r_1 A\ +\ {\sum_{\bf n}}^\prime g^n\ c_{\bf n}\ r_{n_s}\ldots r_{n_2}\,(r_k^{\textrm{gh}} - r_{k-1}^{\textrm{lgt+gh}}r_1)\;A\ ,
	\end{aligned}
	\end{equation}
	where the prime on the sum indicates a restriction on the compositions to $n\ge3$ and $n_1\equiv k>1$,
	\begin{equation}
		{\bf n}=(k,n_2,\ldots,n_s) \qquad\textrm{with}\quad k>1\ .
	\end{equation}
	Hence, 50\%\ of the $2^{n-1}$ compositions (namely those beginning with `1') will not appear.
	We remark that, due to \eqref{fusetrace}, the beginning of any tree diagram contributing to the Nicolai map is either a ghost contribution ($r_k^{\textrm{gh}}$) or the first-order term ($r_1$) followed by a non-invariant contribution ($r_{k-1}^{\textrm{lgt+gh}}$). An invariant contribution $r_i^{\textrm{inv}}$ with $i>1$ can occur only in the second or higher iteration of the coupling flow operator. This reduces the number of contributions considerably in comparison to the non-chiral map.
For higher-order actions of~$\rR_k^{\textrm{inv}}$ the Fierz identity~\eqref{Fierz} can be used to fuse more gamma traces higher up in the tree diagrams, but not for actions of $\rR_k^{\textrm{lgt}}$ or~$\rR_k^{\textrm{gh}}$, because the Lorentz index on their functional variation is attached to $A$ or $\partial$ and not to a gamma matrix. Therefore, multiple gamma traces remain. We have not yet explored these simplifications systematically. 
	
	\noindent\textbf{Chiral Nicolai map to fourth order.\ } We evaluate the traces in \eqref{eq:E} with the identities
	\begin{equation}
		\tfrac14\,\Tr(\gamma_\nu\gamma_\beta\gamma_{\rho\lambda}[\unity{+}\gamma^5])\=-2\eta_{\nu[\rho}\eta_{\lambda]\beta}-\im\,\epsilon_{\nu\beta\rho\lambda}\ ,
		\qquad\qquad\qquad\qquad\qquad\ {}
	\end{equation}
	\begin{equation}
		\begin{aligned}
		\tfrac14\,\Tr\left(\gamma_\nu\gamma_\beta\gamma_\sigma\gamma_\gamma\gamma_{\rho\lambda}[\unity{+}\gamma^5]\right)
		\=-&4(\eta_{\nu[\beta}\eta_{\sigma][\rho}\eta_{\lambda]\gamma}+\eta_{\gamma[\nu}\eta_{\beta][\rho}\eta_{\lambda]\sigma}
		+\eta_{\nu[\rho}\eta_{\lambda][\beta}\eta_{\gamma]\sigma})\\
		-&\im\bigl(\eta_{\nu\beta}\epsilon_{\sigma\gamma\rho\lambda}-\eta_{\nu\sigma}\epsilon_{\beta\gamma\rho\lambda}+\eta_{\beta\sigma}\epsilon_{\nu\gamma\rho\lambda}-2\eta_{\gamma[\rho}\epsilon_{\lambda]\nu\beta\sigma}\bigr)\ ,
		\end{aligned}
	\end{equation}
	where square brackets indicate antisymmetrization of indices.\footnote{with strength one, e.g.~$a_{[\mu}b_{\nu]}=\frac12 (a_\mu b_\nu-a_\nu b_\mu)$.}
	The result for the map is
	\begin{equation}\label{eq:chiral_map}
	\begin{aligned}
	T_gA_\mu \= &A_\mu\ 
	-\ g\ \bigl\{ C_\lambda A_\mu{\times} A^\lambda
 	+ \sfrac{\im}{2} \epsilon_{\mu\alpha\rho\lambda} C^\alpha A^\rho{\times} A^\lambda \bigr\}\\		
	-\, &\sfrac{g^3}{3}\,[A_\mu-C_\mu A\cdot\partial]\, C A_\rho C_\lambda A^\rho{\times} A^\lambda\, 		
	+\, \sfrac{2g^3}{3}\,C^\alpha A_{[\mu} C_{\alpha]} A_\rho C_\lambda A^\rho{\times} A^\lambda \,
	+\, 4 g^3\,C^\nu A^\alpha C^\beta A_{[\mu} C_\nu A_\alpha {\times} A_{\beta]}\\
	-\,&\sfrac{\im g^3}{6}\epsilon_{\nu\sigma\rho\lambda} [A_\mu-C_\mu A\cdot\partial] C A^\nu C^\sigma A^\rho{\times} A^\lambda\,
	+\, \sfrac{\im g^3}{3}\epsilon_{\nu\sigma\rho\lambda} C^\alpha A_{[\mu} C_{\alpha]} A^\nu C^\sigma A^\rho{\times} A^\lambda \\
	-\, &\sfrac{\im g^3}{3}\epsilon_{\mu\nu\alpha\beta}C^\alpha A^\nu C^\beta A_\rho C_\lambda A^\rho{\times} A^\lambda \
	+\ g^4\ T_gA_\mu\bigr|_{{\cal O}(g^4)}\ +\ \mathcal{O}(g^5)\ ,
	\end{aligned}
	\end{equation}
	where in the third order we have two topologies (i.e.~implicit color and position structures), e.g.
	\begin{equation}
		(A_\mu C A_\rho C_\lambda A^\rho{\times} A^\lambda)^a(x)\ \equiv\ 
		f^{abc}f^{cde}f^{efg}\int_{\mathrlap{y_1,y_2}}\quad A_\mu^b(x) C(x{-}y_1)A_\rho^d(y_1)C_\lambda(y_1{-}y_2)A^{f\;\rho}(y_2)A^{g\;\lambda}(y_2)
	\end{equation}
	and
	\begin{equation}
		\begin{aligned}
		&(C_\alpha A_\mu C^\alpha A_\rho C_\lambda A^\rho{\times} A^\lambda)^a(x)\ \equiv\ 
		f^{abc}f^{cde}f^{efg} \smash{\int_{\mathrlap{y_1,y_2,y_3}}}\quad C_\alpha(x{-}y_1)A_\mu^b(y_1) C^\alpha(y_1{-}y_2) \!\!\qquad\qquad\qquad\qquad{} \\
		&\pushright{\cdot\ A_\rho^d(y_2)C_\lambda(y_2{-}y_3)A^{f\;\rho}(y_3)A^{g\;\lambda}(y_3)\ . }
		\end{aligned}
	\end{equation}
	We note that the first branched tree, which the non-chiral map generates from $\rR_1^3\,A$ at this order, is absent here.
	For the fourth order, we introduce the following abbreviations for the various topologies,
	\begin{equation}
		\begin{aligned}
		(G^{1\mathrm{A}}_{\mu\nu\alpha\beta\gamma\delta\sigma\rho\lambda})^a(x)
		&\=f^{abc}f^{cde}f^{efg}f^{ghi} \smash{\int_{\mathrlap{y_1\ldots y_4}}} \quad C_\mu(x{-}y_1)A_\nu^b(y_1) C_\alpha(y_1{-}y_2)A_\beta^d(y_2)C_\gamma(y_2{-}y_3)\\
		&\pushright{\cdot\ A_\delta^f(y_3)C_\sigma(y_3{-}y_4)A^h_\rho(y_4)A^i_\lambda(y_4)\ ,}\\[4pt]
		(G^{1\mathrm{B}}_{\mu\nu\gamma\delta\sigma\rho\lambda})^a(x)
		&\=f^{abc}f^{cde}f^{efg}f^{ghi} \smash{\int_{\mathrlap{y_1\ldots y_4}}} \quad [A^b_\mu(x)\delta(x{-}y_1)-C_\mu(x{-}y_1)A^b(y_1)\cdot\partial]C(y_1{-}y_2)\\
		&\pushright{\cdot\ A_\nu^d(y_2) C_\gamma(y_2{-}y_3)A_\delta^f(y_3)C_\sigma(y_3{-}y_4)A^h_\rho(y_4)A^i_\lambda(y_4)\ ,}\\[4pt]
		(G^{1\mathrm{C}}_{\mu\nu\gamma\delta\sigma\rho\lambda})^a(x)
		&\=f^{abc}f^{cde}f^{efg}f^{ghi} \smash{\int_{\mathrlap{y_1\ldots y_4}}} \quad C_\mu(x{-}y_1)A_\nu^b(y_1)[A^d_\gamma(y_1)\delta(y_1{-}y_2)-C_\gamma(y_1{-}y_2)A^d(y_2)\cdot\partial]\\
		&\pushright{\cdot\ C(y_2{-}y_3)A_\delta^f(y_3)C_\sigma(y_3{-}y_4)A^h_\rho(y_4)A^i_\lambda(y_4)\ ,}\\[4pt]
		(G^{2\mathrm{A}}_{\mu\nu\alpha\beta\gamma\delta\sigma\rho\lambda})^a(x)
		&\=f^{abc}f^{cde}f^{dfg}f^{ehi} \smash{\int_{\mathrlap{y_1\ldots y_4}}} \quad C_\mu(x{-}y_1)A_\nu^b(y_1) C_\alpha(y_1{-}y_2)C_\beta(y_2{-}y_3)A_\gamma^f(y_3)A_\delta^g(y_3)\\
		&\pushright{\cdot\ C_\sigma(y_2{-}y_4)A_\rho^h(y_4)A_\lambda^i(y_4)\ ,}\\[4pt]
		(G^{2\mathrm{B}}_{\mu\beta\gamma\delta\sigma\rho\lambda})^a(x)
		&\=f^{abc}f^{cde}f^{dfg}f^{ehi} \smash{\int_{\mathrlap{y_1\ldots y_4}}} \quad [A^b_\mu(x)\delta(x{-}y_1)-C_\mu(x{-}y_1)A^b(y_1)\cdot\partial]C(y_1{-}y_2)\\
		&\pushright{\cdot\ C_\beta(y_2{-}y_3)A_\gamma^f(y_3)A_\delta^g(y_3)C_\sigma(y_2{-}y_4)A_\rho^h(y_4)A_\lambda^i(y_4)\ ,}\\[4pt]
		(G^{3}_{\mu\nu\alpha\beta\gamma\delta\sigma\rho\lambda})^a(x)
		&\=f^{abc}f^{bde}f^{cfg}f^{ghi} \smash{\int_{\mathrlap{y_1\ldots y_4}}} \quad C_\mu(x{-}y_1)C_\nu(y_1{-}y_2)A_\alpha^d(y_2)A_\beta^e(y_2) C_\gamma(y_1{-}y_3)A_\delta^f(y_3)\\
		&\pushright{\cdot\ C_\sigma(y_3{-}y_4)A^h_\rho(y_4)A^i_\lambda(y_4)\ ,}
		\end{aligned}
	\end{equation}
	which allow us to express the fourth order compactly as (we write all indices downstairs for clarity but pairs of indices are still contracted)
	\begin{equation}
		\begin{aligned}
			T_gA_\mu\bigr|_{{\cal O}(g^4)}\=
			-&G^{1\mathrm{A}}_{\nu[\mu\nu]\beta[\beta\rho]\lambda\rho\lambda}
			-G^{1\mathrm{A}}_{\nu[\mu\nu][\beta\rho]\lambda\beta\rho\lambda}
			-G^{1\mathrm{A}}_{\nu[\mu\nu]\rho[\lambda|\beta|\beta]\rho\lambda}
			-\sfrac13G^{1\mathrm{A}}_{\nu[\mu\nu][\beta\rho]\beta\lambda\rho\lambda}
			+\sfrac12G^{1\mathrm{A}}_{\alpha[\mu\alpha]\nu\sigma\rho[\nu\sigma\rho]}\\
			-&3G^{1\mathrm{A}}_{\nu\sigma\epsilon\delta\delta[\mu\nu\sigma\epsilon]}
			+4G^{1\mathrm{A}}_{\nu\sigma\epsilon\delta[\mu|\delta|\nu\sigma\epsilon]}
			-4G^{1\mathrm{A}}_{\nu\sigma\epsilon[\mu|\delta\delta|\nu\sigma\epsilon]}
			-8G^{1\mathrm{A}}_{\nu\sigma\epsilon[\mu\nu\sigma|\delta\delta|\epsilon]}\\
			-&\sfrac{\im}{2}\epsilon_{\mu\nu\alpha\beta}(
			G^{1\mathrm{A}}_{\nu\alpha\beta\sigma[\sigma\rho]\lambda\rho\lambda}
			+G^{1\mathrm{A}}_{\nu\alpha\beta[\sigma\rho]\lambda\sigma\rho\lambda}
			+G^{1\mathrm{A}}_{\nu\alpha\beta\rho[\lambda|\sigma|\sigma]\rho\lambda}
			+\sfrac13G^{1\mathrm{A}}_{\nu\alpha\beta[\rho\lambda]\rho\gamma\lambda\gamma}
			-\sfrac12G^{1\mathrm{A}}_{\nu\alpha\beta\sigma\delta\rho[\sigma\delta\rho]})\\
			-&\sfrac{\im}{12}\epsilon_{\alpha\beta\rho\lambda}(
			3G^{1\mathrm{A}}_{\nu[\mu\nu]\delta\delta\alpha\beta\rho\lambda}
			-4G^{1\mathrm{A}}_{\nu[\mu\nu][\delta\alpha]\delta\beta\rho\lambda}
			+4G^{1\mathrm{A}}_{\nu[\mu\nu]\alpha\beta\rho\delta\delta\lambda})\\
			+&\sfrac12G^{1\mathrm{B}}_{\mu\beta\beta\rho\lambda\rho\lambda}
			-\sfrac14G^{1\mathrm{B}}_{\mu\beta\rho\beta\lambda\rho\lambda}
			+\sfrac12G^{1\mathrm{B}}_{\mu[\beta\rho]\lambda\beta\rho\lambda}
			+\sfrac12 G^{1\mathrm{B}}_{\mu\rho[\lambda|\beta|\beta]\lambda\rho\lambda}\\
			-&\sfrac16G^{1\mathrm{B}}_{\mu[\beta\rho]\beta\lambda\rho\lambda}
			-\sfrac14G^{1\mathrm{B}}_{\mu\nu\rho\lambda[\nu\rho\lambda]}
			+\sfrac{\im}{12}\epsilon_{\alpha\beta\rho\lambda}(
			3G^{1\mathrm{B}}_{\mu\nu\nu\alpha\beta\rho\lambda}
			+4G^{1\mathrm{B}}_{\mu[\alpha\nu]\nu\beta\rho\lambda}
			+2G^{1\mathrm{B}}_{\mu\alpha\beta\rho\nu\nu\lambda})\\
			-&\sfrac12G^{1\mathrm{C}}_{\lambda[\lambda\mu]\sigma\delta\sigma\delta}
			+3G^{1\mathrm{C}}_{\nu\rho\lambda[\mu\nu\rho\lambda]}
			+\sfrac{\im}{4}\epsilon_{\mu\nu\rho\lambda}G^{1\mathrm{C}}_{\nu\rho\lambda\sigma\delta\sigma\delta}
			+\sfrac{\im}{4}\epsilon_{\sigma\delta\epsilon\gamma}G^{1\mathrm{C}}_{\lambda[\mu\lambda]\sigma\delta\epsilon\gamma}\\
			-&2G^{2\mathrm{A}}_{\nu\alpha\beta[\mu\nu\alpha|\gamma|\beta]\gamma}
			+\sfrac{\im}{12}\epsilon_{\nu\rho\lambda\delta}G^{2\mathrm{B}}_{\mu\alpha\nu\alpha\rho\lambda\delta}\\
			-&\sfrac16 G^3_{\alpha\gamma[\mu|\gamma|\alpha]\rho\lambda\rho\lambda}
			-G^3_{\nu\gamma\alpha\gamma\beta[\mu\nu\alpha\beta]}
			-\sfrac14G^3_{\alpha[\mu\alpha\beta]\beta\rho\lambda\rho\lambda}
			+\sfrac12G^3_{\alpha[\sigma\rho\lambda|\alpha|\mu]\sigma\rho\lambda}
			-\sfrac12G^3_{\alpha[\sigma\rho\lambda|\mu|\alpha]\sigma\rho\lambda}\\
			+&\sfrac{\im}{12}\epsilon_{\mu\nu\alpha\beta}G^3_{\alpha\gamma\nu\gamma\beta\rho\lambda\rho\lambda}
			-\sfrac{\im}{12}\epsilon_{\nu\sigma\rho\lambda}G^3_{\alpha\gamma[\mu|\gamma|\alpha]\nu\sigma\rho\lambda}\\
			-&\sfrac{\im}{12}\epsilon_{[\mu|\nu\alpha\beta}G^3_{\gamma\nu\alpha\beta|\gamma]\rho\lambda\rho\lambda}
			-\sfrac{\im}{8}\epsilon_{\nu\sigma\rho\lambda}G^3_{\alpha[\mu\alpha\gamma]\gamma\nu\sigma\rho\lambda}\ .
		\end{aligned}
	\end{equation}
	\begin{wraptable}{r}{6.0cm}
	\vspace{-0.4cm}
	{\small
	\hspace{0.2cm}
	\begin{tabular}{| l | c c c c |}
	\hline
	order & \ 1\  & \ 2 & \ 3 & 4 \\
	\hline
	non-chiral map & \ 1\ & \ 3 & \ 34 & 380 \\
	chiral map & \ 2\ & \ 0 & \ 21 & 224 \\
	\hline
	\end{tabular}
	}
	\vspace{-0.6cm}
	\end{wraptable}
	Here, indices between vertical lines are omitted from antisymmetrization. Taking into account
        all the antisymmetrizations of indices while respecting the symmetries of the various topologies,
        we count here the number of terms in the first four orders for the non-chiral map~\cite{MN} versus the chiral map~\eqref{eq:chiral_map}.
        There does not seem to be a huge difference between the two formulations,
        but the epsilon symbol generated in the chiral map allows one to combine the antisymmetrization of many terms.

	The tests for the map to the third order \eqref{eq:chiral_map} are performed in the Appendix. These are the gauge condition
	\begin{equation}\label{eq:gauge_cond}
		\partial^\mu (T_g A)_\mu\=\partial^\mu A_\mu \= 0\ ,
	\end{equation}
	on the Landau-gauge hypersurface, the free-action condition
	\begin{equation}\label{eq:fa_cond}
		S_0[T_gA]\=S_g[A]\ ,
	\end{equation}
	and the determinant-matching condition
	\begin{equation}\label{eq:det_cond}
		\det\bigl(\sfrac{\delta A'}{\delta A}\bigr)\= \MSS\ \FP\ ,
	\end{equation}
	where $\MSS$ is the Matthews--Salam--Seiler determinant (technically a Pfaffian for Majorana fermions), and $\FP$ is the Faddeev--Popov determinant, see e.g.~\cite{ALMNPP}. 
	While the first two conditions are straightforward, the determinant matching is more involved to show. One requires the Jacobi identity in color space and the Schouten identity
	\begin{equation}\label{eq:Schouten}
		\eta_{\alpha\beta}\epsilon_{\mu\nu\rho\lambda}+\eta_{\alpha\mu}\epsilon_{\nu\rho\lambda\beta}+\eta_{\alpha\nu}\epsilon_{\rho\lambda\beta\mu}+
		\eta_{\alpha\rho}\epsilon_{\lambda\beta\mu\nu}+\eta_{\alpha\lambda}\epsilon_{\beta\mu\nu\rho}\=0\ ,
	\end{equation}
	which explicitly makes use of $D{=}4$ spacetime dimensions.
	
	\noindent
	{\bf Conclusions.\ }
	We have exploited the option of adding a topological theta term to the super Yang--Mills action in four spacetime dimensions with the aim of simplifying its Nicolai map. If it were not for the ghost sector induced by gauge fixing, the perturbative expansion of the Nicolai map would collapse to a linear $O(g)$ plus a longitudinal $O(g^3)$ expression for the BPS choice of the theta angle, due to a Fierz identity. We call this choice the `chiral Nicolai map'. The ghost sector still renders the map nontrivial, but it is a lot simpler than for vanishing or generic theta angles. For example, it vanishes in second order in the gauge coupling, and the antisymmetrizations among Lorentz indices are more manifest. Although nonperturbatively the BPS choice is admissible only for Euclidean signature, in Minkowski space it just restricts us to the topologically trivial sector. Therefore, perturbative quantum correlators may be computed using the chiral Nicolai map with less effort than previously. To this end, we have written out this map to fourth order in the Landau gauge. In addition, all consistency checks have been verified to the third order, which required several algebraic conspiracies.
	
	The existence of the chiral Nicolai map nurtures the hope that further magical simplifications can occur for the maximally supersymmetric ${\cal N}{=}\,4$ Yang--Mills theory in four dimensions. The SU(4) R-symmetry of this theory allows for even more flexibility in the Nicolai map~\cite{R}. Furthermore, the obstruction to a linear Nicolai map coming from the gauge fixing may perhaps be alleviated by choosing a manifestly supersymmetric gauge fixing, i.e.~a supersymmetric generalization of the Landau gauge (see, e.g.~\cite{CGGJMSV}). We hope to come back to this option of a most simple super Yang--Mills Nicolai map soon.
	
	\noindent
	{\bf Acknowledgment.\ } 
	M.R.~is supported by a PhD grant of the German Academic Scholarship Foundation.
	
	\bigskip
	\appendix
	\noindent\textbf{Appendix (Tests to third order).\ } \\[4pt]
	\noindent\textit{Gauge condition \eqref{eq:gauge_cond}:} The first order is easy to check using symmetry. In the third order, we can remove most terms immediately by symmetry arguments (e.g.~$C_{\mu\alpha}\epsilon^{\mu\alpha\rho\lambda}=0$) and the ghost contributions get projected out by
	\begin{equation}
		\partial^\mu[A_\mu-C_\mu A\cdot\partial]\ldots \= [A\cdot\partial-A\cdot\partial]\ldots \=0\ .
	\end{equation}
	
	\noindent\textit{Free-action condition \eqref{eq:fa_cond}:}
	The free-action condition at first order is
	\begin{equation}
		-\intdx\ A_\mu\Box T_gA^{\mu}|_{O(g)}\ \stackrel{!}{=}\ \intdx\ \partial_\mu A_\nu (A^\mu {\times} A^\nu)\ ,
	\end{equation}
	which is easy to check. The contribution from the $\epsilon_{\mu\alpha\rho\lambda}$ only gives a total derivative. At second order, the condition is
	\begin{equation}\label{eq:fa_cond_second_order}
		-\sfrac12 \intdx\ T_gA_{\mu}|_{O(g)}\Box T_gA^{\mu}|_{O(g)}\ \stackrel{!}{=}\ \sfrac{1}{4}\intdx\ (A_\mu{\times} A_\nu) (A^\mu {\times} A^\nu)\ .
	\end{equation}
	On the left-hand side, there are four terms: The mixed terms proportional to one $\epsilon$ symbol cancel each other. The term proportional to two $\epsilon$ symbols can be written as three terms using the identity 
	\begin{equation}
		\epsilon_{\mu\nu\alpha\beta}\epsilon^{\mu\sigma\rho\lambda}
		\=-\delta\indices{_\nu^\sigma}\delta_{\alpha\beta}^{\rho\lambda}+\delta\indices{_\alpha^\sigma}\delta_{\nu\beta}^{\rho\lambda}-\delta\indices{_\beta^\sigma}\delta_{\nu\alpha}^{\rho\lambda}\ .
	\end{equation}
	Of these, the first one gives the desired term on the right-hand side of \eqref{eq:fa_cond_second_order}, while the two others combine to cancel the remaining term on the left-hand side of \eqref{eq:fa_cond_second_order} without any $\epsilon$ symbols. At third order, we need to show that
	\begin{equation}
		\intdx\ A_\mu \Box T_gA^{\mu}|_{O(g^3)} \ \stackrel{!}{=}\ 0\ .
	\end{equation}
	The ghost contributions vanish due to $\partialA=0$ and symmetry. Further, contracting the last term in the second line of \eqref{eq:chiral_map} with $A_\mu \Box$, one finds that
	\begin{equation}
		(\partial^\nu A^\mu{\times} A^\alpha) C^\beta A_{[\mu} C_\nu A_\alpha{\times} A_{\beta]}
		\=-\sfrac12(A^\mu{\times} A^\alpha) C^{\nu\beta} A_{[\mu} C_\nu A_\alpha{\times} A_{\beta]}\=0\ ,
	\end{equation}
	after integration by parts. With the same integration by parts, the contribution from the last term in the third order of \eqref{eq:chiral_map} vanishes. The remaining four terms cancel pairwise.
	
	\noindent\textit{Determinant matching \eqref{eq:det_cond}:}
	The first order is trivial, since the right-hand side starts at $g^2$ and the left-hand side vanishes due to $f^{aac}=0$. At second order, we need to show that
	\begin{equation}\label{eq:detm_2}
		-\sfrac12 \tr\Bigl[\sfrac{\delta A'}{\delta A}\bigr|_{\mathcal{O}(g)}\sfrac{\delta A'}{\delta A}\bigr|_{\mathcal{O}(g)}\Bigr]\ \stackrel{!}{=}\ -\sfrac{1}{2}g^2[5\;\tr(C_\mu A^\mu C_\nu A^\nu)-2\;\tr(C_\mu A_\nu C^\mu A^\nu)]\ ,
	\end{equation}
	where the traces are over color and position (and Lorentz indices on the left-hand side).
	For the second-order computation, we find
	\begin{equation}
	\begin{aligned}
	\tr\Bigl[\sfrac{\delta A'}{\delta A}\bigr|_{\mathcal{O}(g)}\sfrac{\delta A'}{\delta A}\bigr|_{\mathcal{O}(g)}\Bigr]
	\=&(D{-}1)\; \tr(C_\mu A^\mu C_\nu A^\nu)-\epsilon_{\mu\nu\alpha\beta}\epsilon^{\nu\mu\rho\lambda}\;\tr(C^\alpha A^\beta C_\rho A_\lambda)+2\im\epsilon_{\mu\nu\alpha\beta}\;\tr(C^\mu A^\nu C^\alpha A^\beta)\\
	&(D{+}1)\; \tr(C_\mu A^\mu C_\nu A^\nu)-2\;\tr(C_\mu A_\nu C^\mu A^\nu)+2\im\epsilon_{\mu\nu\alpha\beta}\;\tr(C^\mu A^\nu C^\alpha A^\beta)\ ,
	\end{aligned}
	\end{equation}
	where we used 
	\begin{equation}
	\epsilon_{\mu\nu\alpha\beta}\epsilon^{\nu\mu\rho\lambda}\=2\delta\indices{_\alpha^\rho}\delta\indices{_\beta^\lambda}-2\delta\indices{_\alpha^\lambda}\delta\indices{_\beta^\rho}\ .
	\end{equation}
	With $D{=}4$, this gives exactly what we need in \eqref{eq:detm_2}, but we have one term remaining. However, with $C^\mu(x{-}y)=-C^\mu(y{-}x)$, we get
	\begin{equation}
	\epsilon_{\mu\nu\alpha\beta}\;\tr(C^\mu A^\nu C^\alpha A^\beta)\=\epsilon_{\mu\nu\alpha\beta}\;\tr(C^\alpha A^\nu C^\mu A^\beta)\=0\ ,
	\end{equation}
	so that the condition is indeed satisfied. At third order, the condition is
	\begin{equation}\label{eq:detm_3}
		\begin{aligned}
			\tr\Bigl[\sfrac{\delta A'}{\delta A}\bigr|_{\mathcal{O}(g^3)}\Bigr]+\sfrac13 \tr\Bigl[\sfrac{\delta A'}{\delta A}\bigr|_{\mathcal{O}(g)}\sfrac{\delta A'}{\delta A}\bigr|_{\mathcal{O}(g)}\sfrac{\delta A'}{\delta A}\bigr|_{\mathcal{O}(g)}\Bigr]\ \stackrel{!}{=}\ 
			&+\ 4\  \textcolor{red}{\tr(C_\mu A^\mu C_\rho A_\lambda C^\rho A^\lambda)}\ -\ 
			\sfrac{5}{3}\  \textcolor{blue}{\tr(C_\mu A_\rho C^\rho A_\lambda C^\lambda A^\mu)}\\
			&-\ 2\  \textcolor{orange}{\tr(C_\mu A_\rho C^\rho A^\mu C_\lambda A^\lambda)}\ +\  
			\sfrac{2}{3}\  \textcolor{OliveGreen}{\tr(C_\mu A_\rho C_\lambda A^\mu C^\rho A^\lambda)}\\
			&-\ 2\  \textcolor{violet}{\tr(C_\mu A_\rho C_\lambda A^\mu C^\lambda A^\rho)}\ ,
		\end{aligned}
	\end{equation}
	where we use the same color-coding as in \cite{ALMNPP}.\footnote{with $r{=}4$ spinor degrees of freedom in four spacetime dimensions, 
	and we differ by an overall minus sign on the right-hand side of (3.16) in \cite{ALMNPP} compared to \eqref{eq:detm_3} due to different color ordering.}
	First we compute
	\begin{equation}
	\begin{aligned}
	\sfrac13 \tr\Bigl[\sfrac{\delta A'}{\delta A}\bigr|_{\mathcal{O}(g)}\sfrac{\delta A'}{\delta A}\bigr|_{\mathcal{O}(g)}\sfrac{\delta A'}{\delta A}\bigr|_{\mathcal{O}(g)}\Bigr]
	\=\sfrac13\tr(C^\rho A_\lambda C^\alpha A_\beta C^\sigma A_\delta)\ 
	\bigl[-\,&\delta_{\mu\rho}^{\nu\lambda}\delta_{\nu\alpha}^{\gamma\beta}\delta_{\gamma\sigma}^{\mu\delta}-\im\epsilon\indices{_\mu^\nu_\rho^\lambda}\epsilon\indices{_\nu^\gamma_\alpha^\beta}\epsilon\indices{_\gamma^\mu_\sigma^\delta}\\
	+\,&3\delta_{\mu\rho}^{\nu\lambda}\epsilon\indices{_\nu^\gamma_\alpha^\beta}\epsilon\indices{_\gamma^\mu_\sigma^\delta}+3\im\delta_{\mu\rho}^{\nu\lambda}\delta_{\nu\alpha}^{\gamma\beta}\epsilon\indices{_\gamma^\mu_\sigma^\delta}\bigr]\ ,
		\end{aligned}
	\end{equation}
	for which the various terms give
	\begin{equation}\label{eq:tr_g3_1}
	\begin{aligned}
	-\sfrac13\delta_{\mu\rho}^{\nu\lambda}\delta_{\nu\alpha}^{\gamma\beta}\delta_{\gamma\sigma}^{\mu\delta}\quad
	&\longrightarrow\quad\sfrac{3-D}{3}\ \textcolor{blue}{\tr(C_\mu A_\rho C^\rho A_\lambda C^\lambda A^\mu)}\ -\ 
	\textcolor{orange}{\tr(C_\mu A_\rho C^\rho A^\mu C_\lambda A^\lambda)}\\ 
	&\hspace{5.0cm} +\, \sfrac13\ \textcolor{OliveGreen}{\tr(C_\mu A_\rho C_\lambda A^\mu C^\rho A^\lambda)}\ , \\[4pt]
	-\sfrac13\im\epsilon\indices{_\mu^\nu_\rho^\lambda}\epsilon\indices{_\nu^\gamma_\alpha^\beta}\epsilon\indices{_\gamma^\mu_\sigma^\delta}\quad
	&\longrightarrow\quad \sfrac{\im}{3}\epsilon_{\mu\nu\rho\lambda}\bigl\{\tr(C^\mu A^\alpha C^\nu A_\alpha C^\rho A^\lambda)+\tr(C^\alpha A^\mu C_\alpha A^\nu C^\rho A^\lambda)\\
	&\hspace{1.9cm} -\tr(C^\mu A^\alpha C_\alpha A^\nu C^\rho A^\lambda)-\tr(C^\alpha A^\mu C^\nu A_\alpha C^\rho A^\lambda)\bigr\}\ ,\\[4pt]
	\delta_{\mu\rho}^{\nu\lambda}\epsilon\indices{_\nu^\gamma_\alpha^\beta}\epsilon\indices{_\gamma^\mu_\sigma^\delta}\quad
	&\longrightarrow\quad -\ \textcolor{blue}{\tr(C_\mu A_\rho C^\rho A_\lambda C^\lambda A^\mu)}\ +\ 
	(3{-}D)\ \textcolor{orange}{\tr(C_\mu A_\rho C^\rho A^\mu C_\lambda A^\lambda)}\\
	&\hspace{1.1cm} +\, (D{-}1)\ \textcolor{red}{\tr(C_\mu A^\mu C_\rho A_\lambda C^\rho A^\lambda)}\ -\ 
	\textcolor{violet}{\tr(C_\mu A_\rho C_\lambda A^\mu C^\lambda A^\rho)}\ ,\\[4pt]
	\im\delta_{\mu\rho}^{\nu\lambda}\delta_{\nu\alpha}^{\gamma\beta}\epsilon\indices{_\gamma^\mu_\sigma^\delta}\quad
	&\longrightarrow\quad -\im\epsilon_{\mu\nu\rho\lambda}\bigl\{2\tr(C^\alpha A_\alpha C^\mu A^\nu C^\rho A^\lambda)+\tr(C^\alpha A^\mu C^\nu A_\alpha C^\rho A^\lambda)\bigr\}\ .
	\end{aligned}
	\end{equation}
	For the first term on the left-hand side of \eqref{eq:detm_3}, we list the contributions of the third order of \eqref{eq:chiral_map} in one line per term:\footnote{separating the contributions from the antisymmetrization $[\mu\,\alpha]$ in two lines (3rd+4th and 8th+9th) and collecting the contributions from the antisymmetrization $[\mu\,\nu\,\alpha\,\beta]$ in one line (5th).}
	\begin{equation}\label{eq:tr_g3_2}
	\begin{aligned}
	\tr\Bigl[\sfrac{\delta A'}{\delta A}\bigr|_{\mathcal{O}(g^3)}\Bigr] \=
	\textcolor{gray}{-}\ &\textcolor{gray}{\sfrac{N}{3} A_\mu (C) C_\lambda A^{\mu}\times A^{\lambda}\ -\ 
	\sfrac23 \tr(A^\mu C A_{[\mu}C_{\lambda]}A^\lambda)}\\
	\textcolor{cyan}{+}\ &\textcolor{cyan}{\sfrac{N}{3} A^\alpha (C_\mu C_\alpha)C_\lambda A^\mu\times A^\lambda}\ +\ 
	\sfrac13\ \textcolor{orange}{\tr(C_\mu A_\rho C^\rho A^\mu C_\lambda A^\lambda)}\ -\ 
	\sfrac13\ \textcolor{red}{\tr(C_\mu A^\mu C_\rho A_\lambda C^\rho A^\lambda)}\\
	\textcolor{gray}{+}\ &\textcolor{gray}{\sfrac{N}{3} A_\mu (C^\alpha C_\alpha)C_\lambda A^\mu\times A^\lambda}\ -\ 
	\sfrac13\ \textcolor{violet}{\tr(C_\mu A_\rho C_\lambda A^\mu C^\lambda A^\rho)}\ +\ 
	\sfrac13\ \textcolor{red}{\tr(C_\mu A^\mu C_\rho A_\lambda C^\rho A^\lambda)}\\
	\textcolor{cyan}{-}\ &\textcolor{cyan}{\sfrac{N}{3} A^\alpha (C_\mu C_\alpha)C_\lambda A^\mu\times A^\lambda}\ -\ 
	\sfrac13\ \textcolor{blue}{\tr(C_\mu A_\rho C^\rho A_\lambda C^\lambda A^\mu)}\ +\ 
	\sfrac13\ \textcolor{red}{\tr(C_\mu A^\mu C_\rho A_\lambda C^\rho A^\lambda)}\\
	-\ &\sfrac13\ \textcolor{orange}{\tr(C_\mu A_\rho C^\rho A^\mu C_\lambda A^\lambda)}\ -\ 
	\sfrac23\ \textcolor{violet}{\tr(C_\mu A_\rho C_\lambda A^\mu C^\lambda A^\rho)}\ +\ 
	\sfrac23\ \textcolor{red}{\tr(C_\mu A^\mu C_\rho A_\lambda C^\rho A^\lambda)}\\
	&\pushright{+\ \sfrac13\ \textcolor{OliveGreen}{\tr(C_\mu A_\rho C_\lambda A^\mu C^\rho A^\lambda)}}\\
	\textcolor{gray}{-}\ &\textcolor{gray}{\im \sfrac{N}{6} \epsilon_{\mu\nu\rho\lambda}A^\mu (C) C^\nu A^{\rho}\times A^{\lambda}\ -\ 
	\sfrac{\im}{3}\epsilon_{\mu\nu\rho\lambda}\tr(A^\mu C A^{\nu}C^{\rho}A^\lambda)}\\
	\textcolor{cyan}{+}\ &\textcolor{cyan}{\im \sfrac{N}{6} \epsilon_{\mu\nu\rho\lambda}A_\alpha(C^\alpha C^\mu) C^\nu A^\rho\times A^\lambda}\ +\ 
	\sfrac{\im}{3}\epsilon_{\mu\nu\rho\lambda}\;\tr(C^\mu A^\alpha C_\alpha A^\nu C^\rho A^\lambda)\\
	\textcolor{gray}{+}\ &\textcolor{gray}{\im \sfrac{N}{6} \epsilon_{\mu\nu\rho\lambda}A^\mu (C^\alpha C_\alpha)C^\nu A^\rho\times A^\lambda}\ +\ 
	\sfrac{\im}{3}\epsilon_{\mu\nu\rho\lambda}\;\tr(C_\alpha A^\mu C^\alpha A^\nu C^\rho A^\lambda)\\
	\textcolor{cyan}{-}\ &\textcolor{cyan}{\im \sfrac{N}{6} \epsilon_{\mu\nu\rho\lambda}A_\alpha(C^\alpha C^\mu) C^\nu A^\rho\times A^\lambda}\ -\ 
	\sfrac{\im}{3}\epsilon_{\mu\nu\rho\lambda}\;\tr(C_\alpha A^\alpha C^\mu A^\nu C^\rho A^\lambda)\\
	-\ & \sfrac{\im}{3}\epsilon_{\mu\nu\rho\lambda}\;\tr(C_\alpha A^\alpha C^\mu A^\nu C^\rho A^\lambda)\ -\ 
	\sfrac{\im}{3}\epsilon_{\mu\nu\rho\lambda}\;\tr(C^\mu A^\alpha C^\nu A^\rho C^\lambda A_\alpha)\ ,
	\end{aligned}
	\end{equation}
	where round brackets indicate a loop.
	The gray terms vanish in groups of three in a calculation already performed in 2005.12324 (3.23), making use of the Jacobi identity. The cyan terms cancel pairwise. 
	For the black terms, we first note that we can read the traces `backwards', e.g.~
	\begin{equation}
		\tr(C^\alpha A^\beta C^\mu A^\nu C^\rho A^\lambda)\=\tr(C^\alpha A^\lambda C^\rho A^\nu C^\mu A^\beta)\ ,
	\end{equation}
	using $C^\alpha(x{-}y)=-C^{\alpha}(y{-}x)$ and $f^{abc}A^b_\mu=-f^{cba}A^b_\mu$ (giving us six minus signs, hence a plus overall).
	This, together with the cyclicity of the trace and symmetry, gives us
	\begin{equation}\label{eq:simplf}
		\begin{aligned}
			&\epsilon_{\mu\nu\rho\lambda}\tr(C^\mu A^\alpha C^\nu A_\alpha C^\rho A^\lambda)\=\epsilon_{\mu\nu\rho\lambda}\tr(C^\rho A_\alpha C^\nu A^\alpha C^\mu A^\lambda )\=-\epsilon_{\mu\nu\rho\lambda}\tr(C^\mu A^\alpha C^\nu A_\alpha C^\rho A^\lambda)\=0\ ,\\
			&\epsilon_{\mu\nu\rho\lambda}\tr(C^\alpha A^\mu C_\alpha A^\nu C^\rho A^\lambda)\=\epsilon_{\mu\nu\rho\lambda}\tr(C_\alpha A^\mu C^\alpha A^\lambda C^\rho A^\nu)\=-\epsilon_{\mu\nu\rho\lambda}\tr(C^\alpha A^\mu C_\alpha A^\nu C^\rho A^\lambda)\=0\ ,\\
			&\epsilon_{\mu\nu\rho\lambda}\tr(C^\mu A^\alpha C_\alpha A^\nu C^\rho A^\lambda)\=\epsilon_{\mu\nu\rho\lambda}\tr(C_\alpha A^\alpha C^\mu A^\lambda C^\rho A^\nu)\=-\epsilon_{\mu\nu\rho\lambda}\tr(C_\alpha A^\alpha C^\mu A^\nu C^\rho A^\lambda)\ .
		\end{aligned}
	\end{equation}
	Further we can use the Schouten identity \eqref{eq:Schouten}, which implies that
	\begin{equation}\label{eq:Schouten2}
		\begin{aligned}
			0\=&\epsilon_{\mu\nu\rho\lambda}[\tr(C^\alpha A_\alpha C^\mu A^\nu C^\rho A^\lambda)-\tr(C^\alpha A^\mu C_\alpha A^\nu C^\rho A^\lambda)+
			\tr(C^\alpha A^\mu C^\nu A_\alpha C^\rho A^\lambda)\\
			&\hspace{4.18cm}-\tr(C^\alpha A^\mu C^\nu A^\rho C_\alpha A^\lambda)+\tr(C^\alpha A^\mu C^\nu A^\rho C^\lambda A_\alpha)]\\
			\=&2 \epsilon_{\mu\nu\rho\lambda}\tr(C^\alpha A_\alpha C^\mu A^\nu C^\rho A^\lambda)+\epsilon_{\mu\nu\rho\lambda}\tr(C^\alpha A^\mu C^\nu A_\alpha C^\rho A^\lambda)\ .
		\end{aligned}
	\end{equation}
	Applying \eqref{eq:simplf} to the black terms in \eqref{eq:tr_g3_1} and \eqref{eq:tr_g3_2}, many terms drop out, and we are left with
	\begin{equation}
	-\sfrac83 \epsilon_{\mu\nu\rho\lambda}\tr(C^\alpha A_\alpha C^\mu A^\nu C^\rho A^\lambda) -\sfrac43 \epsilon_{\mu\nu\rho\lambda}\tr(C^\alpha A^\mu C^\nu A_\alpha C^\rho A^\lambda)\=0\ ,
	\end{equation}
	which vanishes by \eqref{eq:Schouten2}.
	Lastly, all the colored terms add up to the correct factors needed on the right-hand side of \eqref{eq:detm_3}. 
	This proves the determinant matching condition up to and including third order.

\end{document}